\documentclass[12pt]{article}
\pdfoutput=1
\usepackage{jheppub}
\usepackage{amssymb,amsmath,amstext,amsfonts}
\usepackage{bbold,ulem,bm}
\usepackage{graphics}
\usepackage{mathtools}
\usepackage{hyperref}
\usepackage{color}

\usepackage{graphicx}
\usepackage{mathrsfs}

\def\mpl{M_{\rm p}}

\begin{document}

\title{Conformal solids and holography} 

\author[a]{A.~Esposito,}
\affiliation[a]{Department of Physics, Center for Theoretical Physics, Columbia University, 538W 120th Street, New York, NY, 10027, USA}
\author[b]{S.~Garcia-Saenz,}
\affiliation[b]{Sorbonne Universit\'es, UPMC Univ.\ Paris 6 and CNRS, UMR 7095, Institut d'Astrophysique de Paris, GReCO, 98bis boulevard Arago, 75014 Paris, France}
\author[a]{A.~Nicolis,}
\author[c]{and R.~Penco}
\affiliation[c]{Center for Particle Cosmology, Department of Physics and Astronomy, University of Pennsylvania 209 S. 33rd St., Philadelphia, PA 19104, USA}

\abstract{We argue that a $SO(d)$ magnetic monopole in an asymptotically AdS space-time is dual to a $d$-dimensional strongly coupled system in a solid state. In light of this, it would be remiss of us not to dub such a field configuration {\it solidon}. In the presence of mixed boundary conditions, a solidon spontaneously breaks translations (among many other symmetries) and gives rise to Goldstone excitations on the boundary---the phonons of the solid. We derive the quadratic action for the boundary phonons in the probe limit and show that, when the mixed boundary conditions preserve conformal symmetry, the longitudinal and transverse sound speeds are related to each other as expected from effective field theory arguments. We then include backreaction and calculate the free energy of the solidon for a particular choice of mixed boundary conditions, corresponding to a relevant multi-trace deformation of the boundary theory. We find such free energy to be lower than that of thermal AdS. This suggests that our solidon undergoes a solid-to-liquid first order phase transition by melting into a Schwarzschild-AdS black hole as the temperature is raised.} 


\maketitle


\section{Introduction} \label{sec:intro}

In recent years the AdS/CFT correspondence~\cite{Maldacena:1997re,Witten:1998qj,Gubser:1998bc} has provided a powerful tool for the study of condensed matter systems---see~\cite{Hartnoll:2009sz,Herzog:2009xv,Hartnoll:2016apf} for reviews on the topic. The correspondence relates a weakly coupled theory of gravity on AdS spacetime to a strongly coupled field theory living on its boundary. It therefore allows to compute physical quantities in a non-perturbative regime, where other standard techniques fail.

This approach has already been very successful in identifying and studying the gravity dual of systems such as ordinary fluids, superfluids/superconductors, and so on (see {\it e.g.}~\cite{Bhattacharyya:2008jc,Nickel:2010pr,deBoer:2015ija,Crossley:2015tka,Hartnoll:2008vx,Gubser:2009cg,Horowitz:2009ij,Esposito:2016ria}), as well as analyzing new possible exotic states of matter (see {\it e.g.}~\cite{Policastro:2001yc,Kovtun:2004de,Buchel:2009ge}). In this work, we build on these achievements and provide an explicit example of a conformal solid. 

In order to do that, we draw our inspiration from the effective field theory (EFT) description of solids~\cite{Soper:1976bb,Leutwyler:1996er,Son:2005ak,Dubovsky:2005xd,Endlich:2012pz}. In this language, the low energy description of a solid (as well as other states of matter~\cite{Nicolis:2015sra}) is completely given in terms of its symmetry breaking pattern, which makes it fairly simple to deduce the gravity dual from the holographic dictionary.

In particular, we find that the dual gravitational description of a solid in $d$ spacetime dimensions  is given by a magnetic monopole in an $SO(d)$ Yang--Mills theory coupled to a scalar field in the fundamental representation. We cannot resist the temptation of attaching to such a field configuration the apt moniker of {\it solidon}. 

Solidons have already been extensively studied for their own sake (see {\it e.g.}~\cite{Lugo:1999ai,Lugo:1999fm}), and some indications that they might be dual to a solid already appeared in the literature~\cite{Bolognesi:2010nb,Sutcliffe:2011sr}. Some steps towards the identification of the gravity dual of a solid have also been made in~\cite{Alberte:2015isw,Alberte:2016xja, Alberte:2017cch} using theories of massive gravity. We will come back to these other approaches at the end of our paper.

We first review the EFT for solids in flat space and derive the constraints that conformal symmetry places on the longitudinal and transverse sound speeds. We then discuss the effective theory for a solid on a sphere, whose holographic dual is at first easier to identify compared to  that of a flat solid. 

After introducing our bulk theory, we show that it admits a solidon configuration in the presence of mixed boundary conditions by solving numerically the background equations of motion in the probe limit. We then use the methods introduced in~\cite{deBoer:2015ija} and extended in~\cite{Esposito:2016ria} to calculate explicitly the quadratic action for the boundary phonons. We regard this as a strong evidence for the duality between our solidon configuration and a solid on the boundary. 
Finally, we reintroduce the backreaction to calculate the free energy of the solidon, and discuss the existence of a first order phase transition between the solidon and a Schwarzschild-AdS (SAdS) black hole, which as is well-known is dual to a fluid (see {\it e.g.}~\cite{Bhattacharyya:2008jc,Nickel:2010pr,deBoer:2015ija,Crossley:2015tka}).

\vspace{2em}

\noindent {\it Conventions:} Throughout this paper we will work in units such that $\hbar=c=L=1$ ($L$ being the AdS radius), and will adopt a ``mostly plus'' metric signature. We will denote with $d$ the number of spacetime dimensions of the boundary theory. Moreover, we use capital letters $M,\,N,\,\dots$ for bulk indices, which are always contracted with the AdS metric, and greek letters $\mu,\,\nu,\,\dots$ for the boundary indices, which are instead contracted with the Minkowski metric (for a flat boundary).


\section{Effective theory of solids} \label{sec:EFT}

At distances much larger than the scale of possible microscopic inhomogeneities, such as crystal structure, a solid can be thought of as a continuous medium.  Its mechanical deformations can then be characterized by specifying the positions of its volume elements, whose comoving coordinates we will denote with $\phi^I$, $I=1,\ldots,d-1$. These can be thought as a set of scalar fields, $\phi^I (t, \vec x)$, which associate a volume element to each position $\vec x$ at any time $t$. The internal symmetries that act on them can be derived as follows. For homogeneous solids in equilibrium at a fixed external pressure, we can always choose comoving coordinates that are aligned with the spatial ones, {\it i.e.}
\begin{equation} \label{solid background}
	 \langle \phi^I \rangle_{\rm eq} = \alpha \, x^I,
\end{equation}
with $\alpha$ some constant that depends on the external pressure or equivalently on the degree of compression or stretching of the solid. This configuration however breaks spatial translations. Since the solid we are considering is supposed to be homogeneous (at large scales), there must be an internal shift symmetry acting on the $\phi^I$'s that is also spontaneously broken, and that combined with spatial translations yields an unbroken translation group. The same considerations apply to rotations: if for simplicity we focus on a solid that is isotropic (again, at large scales), then there must be an internal rotational symmetry acting on the $\phi^I$'s to make up for the spontaneous breaking of rotations in \eqref{solid background}.\footnote{One can easily account for possible anisotropies due to the underlying lattice by proceeding along the lines of~\cite{Kang:2015uha}. In this paper we will restrict ourselves to the isotropic case.} We thus postulate that the dynamics of the solid be invariant under
\begin{equation} \label{solid symmetries}
	\phi^ I \to R^I{}_J \phi^J + c^I \, ,
\end{equation}
for any constant vector $c^I$ and $SO(d-1)$ matrix $R^I {}_J$. The field configuration \eqref{solid background} spontaneously breaks the internal symmetries \eqref{solid symmetries} together with spatial translations and rotations (not to mention boosts) down to the diagonal subgroup~\cite{Nicolis:2013lma}. It is this unbroken subgroup that encodes the large scale isotropy and homogeneity of the solid.
Fluctuations of the $\phi$'s around this equilibrium configuration correspond to phonon excitations.

In order to write down an effective action for the fields $\phi^I$, we exploit invariance under the shifts in Eq.~\eqref{solid symmetries} and Lorentz symmetry to conclude that, at lowest order in the derivative expansion, it can depend on the $\phi$'s only through the combination $B^{IJ}\equiv \partial_\mu \phi^I \partial^\mu \phi^J$~\cite{Dubovsky:2005xd}. Moreover, in $d-1$ spatial dimensions there are only $d-1$ independent invariants under internal rotations that can be built out of the symmetric matrix $B^{IJ}$. Defining $X \equiv \text{tr} B$ and $Y_n \equiv \text{tr} (B^n) / X^n$, the low-energy effective action reads:
\begin{equation} \label{S solids}
	S = \int d^d x \, F (X, Y_2, \cdots, Y_{d-1}) \, ,
\end{equation}
where $F$ is an a priori generic function, related to the equation of state of the solid under consideration.

The quadratic action for the excitations about \eqref{solid background} can be obtained by plugging $\phi^I = \alpha(x^I + \pi^I (t, \vec x))$ into the equation above and expanding up to quadratic order in the phonon fields $\pi^i$. Notice that we can now stop differentiating between internal ($I,J,\dots$) indices and spatial ($i,j, \dots$) ones: under the unbroken linear combination of internal and spatial rotations, they transform in the same way. It is convenient to decompose the phonon field into longitudinal and transverse components, {\it i.e.}\ $\vec \pi = \vec \pi_L + \vec \pi_T$, where $\vec \pi_L$ is the gradient of a scalar and $\vec \pi_T$ is divergenceless. Then, the quadratic action for phonons reads
\begin{equation}
	S_{(2)} = - \frac{F_X X}{d-1} \int d^d x \left[ \dot{\vec \pi}^2 - c_L^2 (\partial_i \pi_L ^j)^2- c_T^2  (\partial_i \pi_T ^j)^2\right] .\nonumber 
\end{equation}
where $X$ and $F_X \equiv \partial F / \partial X$ are evaluated on the background \eqref{solid background}, and we have introduced the longitudinal and transverse phonon speeds
\begin{subequations} \label{sound speeds}
\begin{align}
	c_T^2 =& 1 + \frac{d-1}{F_X X} \cdot \Xi\,, \\
	c_L^2 =& 1 + \frac{2 F_{XX} X^2}{(d-1)F_X X} + \frac{2 (d-2)}{F_X X} \cdot \Xi\,, \\
	\Xi \equiv & \sum_{n=2}^{d-1} \frac{\partial F}{\partial Y_n} \frac{n(n-1)}{(d-1)^n} \; .
\end{align}
\end{subequations}

\subsection{Conformal solids} \label{sec:conformal solids}

In this paper, we will be particularly interested in solid states in a conformal field theory. By a solid state in general we mean a state that at long distances can be described by the same degrees of freedom and the same symmetries as we just described, regardless of whether the microscopic structure resembles that of an ordinary solid. To the best of our knowledge, whether CFTs admit solid states in this sense is an open question. Notice that, by scale invariance, if a solid state in a CFT is not homogeneous down to arbitrarily small distances and there is an underlying lattice structure, then all lattice spacings must be allowed. Moreover, like in the case of a fluid state, conformal invariance must be spontaneously broken. This is perhaps most obvious from the non-vanishing of $\langle T^{\mu\nu} \rangle$, which breaks scale invariance. 

Let us thus assume that we have a solid state in a CFT. Conformal symmetry restricts the form of the otherwise arbitrary function $F$ appearing in the effective action. In fact, by varying Eq.~\eqref{S solids} with respect to the metric, we find the stress energy tensor
\begin{equation}
	T_{\mu\nu} = \eta_{\mu\nu} F -2 \frac{\partial F}{\partial X} B^{-1}_{IJ} \partial_\mu \phi^I \partial_\nu \phi^J \, . 
\end{equation}
Imposing now that the trace of $T_{\mu\nu}$ vanish for arbitrary field configurations---as required by conformal symmetry---yields a differential equation for $F$ that completely fixes its $X$-dependence:
\begin{equation}
	F = X^{d/2} f(Y_2, \cdots, Y_{d-1})\,.
\end{equation}

Using this expression, it is easy to show that the sound speeds \eqref{sound speeds} are no longer independent of each other, but instead obey the universal relation
\begin{equation} \label{conformal sound speeds}
	c_L^2 = \frac{1}{d-1} + 2 \frac{d-2}{d-1} c_T^2 \, .
\end{equation}
In what follows, we will treat this relation as the hallmark of a conformal solid, and we will check that indeed our bulk theory in AdS gives rise to phonon excitations on the boundary whose dispersion relations satisfy Eq.~\eqref{conformal sound speeds}. Notice that, in the absence of instabilities or superluminality, the relation above implies that, for a conformal solid
\begin{equation}
	0 \leqslant c_T^2 \leqslant 1/2\,, \qquad\quad  1/(d-1) \leqslant c_L^2 \leqslant 1\,.
\end{equation}
Hence, the longitudinal phonon speed for a conformal solid is always relativistic.

We should also point out that conformal solids realize conformal symmetry non-linearly without the need for a dilaton field. This is because the phonons fields $\pi^i$, which already in generic solids play the role of Goldstones for multiple broken symmetries~\cite{Nicolis:2013lma}, realize non-linearly also dilations and special conformal transformations in the case of  a conformal solid. This is a manifestation of a much more general result: when it comes to spontaneously broken space-time symmetries, there are generically fewer Goldstone modes than broken symmetries~\cite{Low:2001bw}.

\subsection{Solids on a sphere} \label{sec:solsphere}

For reasons that will be discussed in a moment, it will be easier to introduce the holographic dual of our solid when this lives on a sphere rather than on flat space. Notice that we have to consider a sphere much larger than the scale at which the phonon effective field theory breaks down, otherwise there is no physical sense in which we can say to have a solid on a sphere. Related to this, if our solid has a crystalline microscopic structure, we can  disregard the effect of  dislocations due to the curvature of the sphere provided these are distributed somewhat homogeneously. 

Now we come to the main point: to write the EFT for a solid on a sphere, we have to change the {\it internal} symmetries that define our solid. This is because their role is to combine with the spontaneously broken isometries of the underlying space to yield unbroken combinations that correspond to the homogeneity and isotropy of a solid at equilibrium. Since the isometries of a sphere are different than those of flat space, the internal symmetries of the solid EFT must also change in going from one space to the other. This is perhaps surprising at first---after all, ordinary solids come in all possible shapes---but it is in fact associated with a very familiar fact about solids: once formed, say by solidifying a liquid, a solid at equilibrium has a specific shape, which cannot be changed easily. One cannot bend a flat slab of ice so as to create some spherical curvature, even just locally, and hope that that will be an equilibrium configuration. However, freezing water on a spherical surface is clearly doable, and yields a spherical shell of ice. This means that the two systems are described by two different effective field theories: what is an equilibrium configuration for one, is not an equilibrium configuration for the other. An identical phenomenon happens in massive gravity, which in the St\"uckelberg formalism can be thought of as an exotic solid coupled to general relativity \cite{Alberte:2011ah}.

So, on a sphere the internal $ISO(d-1)$ symmetry \eqref{solid symmetries} acting on the comoving coordinates has  to be replaced by an internal $SO(d)$, since this is the isometry group of a $(d-1)$-dimensional sphere. The comoving coordinates can be taken to be $d-1$ angles $\phi^I$ that at equilibrium are aligned with the angles parametrizing the sphere:
\begin{equation} \label{vev sphere}
	\langle \phi^I \rangle_{\rm eq}= \theta^I \, .
\end{equation}
Similarly to solids in flat space, this field configuration spontaneously breaks the isometries of the sphere and the internal $SO(d)$ (as well as boosts, of course) down to the diagonal subgroup. 
There is however one important difference, namely that a homogeneous and isotropic solid in flat space can be compressed, dilated, and even sheared uniformly by varying continuously the parameter $\alpha$ in Eq.\ \eqref{solid background}, or turning $\alpha$ into a matrix. To the contrary, a solid on a sphere cannot be compressed (or dilated) without spoiling homogeneity---which is why there is no free parameter in Eq.\ \eqref{vev sphere}.  

 There are two different but completely equivalent ways of writing down an effective action for the $\phi$'s that is invariant under $SO(d)$.  The first one is to use again the matrix $B^{IJ} = \partial_\mu \phi^I \partial^\mu \phi^J$ (with greek indices now contracted with the spacetime metric on $\mathbb{R} \times S^{d-1}$) and build invariants by contracting $I,J$ indices with a field-space spherical metric:      
\begin{equation}
	g_{IJ} (\phi) = \text{diag} \left(1, \sin^2 \phi_1, \cdots, \prod_{n=1}^{d-2} \sin^2 \phi_n \right) \, .
\end{equation}

The second method---closer in spirit to the holographic implementation discussed in the next section---is to think of the $(d-1)$-sphere as embedded in a $d$-dimensional flat space, and introduce a multiplet $\vec \Phi(x)$ in the fundamental representation of $SO(d)$ that, at equilibrium, acquires an expectation value in the radial direction. The radial unit vector can always be obtained starting from some fiducial unit vector, which we take to be in the $\hat x_d$ direction, and applying a rotation $R (\theta)$ parametrized by the $d-1$ angles~$\theta^i$ on the unit sphere. Thus, the equilibrium configuration of $\vec \Phi$ reads: 
\begin{equation} \label{EFT vev sphere}
	\langle \vec \Phi \rangle_{\rm eq} =  \Phi_0 \, R(\theta) \cdot \hat x_d \, .
\end{equation}
This vev realizes precisely the symmetry reaking pattern of a solid. If we now replace the angles $\theta^i$ with the fields $\phi^I$, we find that traces of products of the matrix $\mathcal{B}^{AB} (\phi) = \partial_\mu \Phi^A \partial^\mu \Phi^B$ are invariant under $SO(d)$, which is non-linearly realized on the $\phi$'s. The effective action for the $\phi$'s then reads
\begin{equation}
	S = \mathcal{R}^{d-1} \int dt \, d\Omega_{d-1} F (X, Y_2, \cdots, Y_{d-2}) \, ,
\end{equation}
where $\mathcal{R}$ is the radius of the sphere, and $X$ and the $Y_n$'s are now built out of $\mathcal{B}^{AB}$. 

Let us now focus on a small patch of the sphere---with size much smaller than the curvature radius---by replacing $\theta^i\to x^i/\mathcal{R}$ with $x^i\ll\mathcal{R}$. An observer on such a patch who has only access to momenta $k \gg 1/ \mathcal{R}$ is unable to probe the global properties of the space. Thus, $1/\mathcal{R}\equiv\alpha$ effectively plays the role of a free parameter, to be eventually determined from boundary conditions. In this limit, we recover the field configuration \eqref{solid background} describing a solid in flat space. We will take a similar limit later on in AdS, when passing from global to Poincar\'e coordinates.


\section{Holographic dual} \label{sec:holo}

In this section we will describe the gravity dual of the EFT presented above. 
We will first set up the bulk theory dual to a solid on a sphere. We should stress however that we regard this just as a particularly transparent way of compactifying our symmetry breaking pattern. We are ultimately interested in studying flat solids and we will do that by taking the Poincar\'{e} patch limit of our bulk theory.


\subsection{Set-up} \label{sec:setup}

A basic entry of the holographic dictionary states that global symmetries on the boundary should be gauged in the bulk. We thus include an $SO(d)$ Yang--Mills sector in the bulk theory in order to reproduce the internal symmetries of our solid on the sphere. In light of what was discussed in Section~\ref{sec:EFT}, we expect the gravity dual of our system to also include scalar fields responsible for the solid's symmetry breaking pattern. We are therefore led to considering the following action in the bulk of AdS$_{d+1}$:
%
\begin{align} \label{eq:Sm}
S &=-\int d^{d+1}x\sqrt{-g}\bigg[ \frac{1}{2}\,D_M\vec\Phi\cdot D^M\vec\Phi+V\big(|\vec\Phi|^2\big)+\frac{1}{8}\,F_{MN}^{AB}F^{AB\,MN}\bigg]+S_\text{bdy}\,,
\end{align}
%
where $\vec\Phi$ is a real scalar field belonging to the fundamental representation of $SO(d)$, with a potential $V\big(|\vec\Phi|^2\big)$. We will assume that such a potential takes the generic form
\begin{align}
V\big(|\vec\Phi|^2\big)=\frac{m^2}{2}|\vec\Phi|^2+\text{interaction terms}\,.
\end{align}
We will be agnostic about the detailed form of the interaction terms, although later on we will place some restrictions on the values of $m^2$. Finally, $S_\text{bdy}$ is a boundary term needed to render the on-shell action finite as well as to fix the variational problem---see below.

For the purposes of studying the spectrum of phonon excitations, we will neglect the backreaction of the matter fields on the geometry and work at zero temperature. Hence, we will consider a purely AdS background metric:\footnote{It is possible that when the backreaction of the metric is included, only some potentials will give solutions compatible with the background~\eqref{eq:metric}, in the limit of small matter fields. We assume here that we are working with such potentials. For a discussion of this point in the case of superfluids, see for instance~\cite{Esposito:2016ria} and references therein.}
\begin{align} \label{eq:metric}
ds^2=-\left(1+\rho^2\right)d\tau^2+\frac{d\rho^2}{1+\rho^2}+\rho^2 d\Omega_{d-1}^2\,.
\end{align}
In these coordinates the center of AdS is located at $\rho=0$ while the boundary is at $\rho=\infty$. 

We find it more convenient to work in a notation where $A,\,B,\,C, \dotsc =1,\dots,d$ are all indices in the fundamental representation of $SO(d)$. The real generators are then given by
\begin{align}
(T^{AB})_{IJ}=\delta^A_J\delta^B_I-\delta^A_I\delta^B_J\,.
\end{align}
In this formalism, the group algebra is 
\begin{align}
\big[T^{AB},T^{CD}\big]=\frac{1}{2}f^{ABCDEF}T^{EF}\,,
\end{align}
with structure constants
\begin{align}
f^{ABCDEF}&=\delta^{AC}\delta^{BE}\delta^{DF}+\delta^{AE}\delta^{BD}\delta^{CF}  \\
&-\delta^{AD}\delta^{BE}\delta^{CF} - \delta^{AE}\delta^{BC}\delta^{DF} -(E\leftrightarrow F)\,. \notag 
\end{align}

Inspired by our discussion in the previous section, we choose the following ansatz for the background fields:
\begin{align} \label{spherical ansatz 1}
\vec\Phi=\varphi(\rho)\hat \rho,\quad\quad A_M^{AB}=\rho\,\psi(\rho)\partial_M\hat \rho^I(T^{AB})_{IJ} \hat\rho^J,
\end{align}
where $\hat\rho$ is the unit vector pointing to the AdS boundary. The ansatz for the scalar field is clearly motivated by Eq.\ \eqref{EFT vev sphere}, with the vev $\vec \Phi$ now a function of the holographic coordinate. The ansatz for the gauge field is instead a straightforward generalization of the one used for the 't~Hooft-Polyakov monopole solution~\cite{Weinberg:2012pjx}.  The equations of motion for the field profiles are:
%
\begin{subequations} \label{eqphipsi}
\begin{align}
\varphi^{\prime\prime}+\frac{(d-1)+(d+1)\rho^2}{\rho(1+\rho^2)}\varphi^\prime-\frac{2V^\prime(\varphi^2)}{1+\rho^2}\varphi-(d-1)\frac{\big(1+q\rho\,\psi\big)^2}{\rho^2(1+\rho^2)}\varphi&=0\,, \label{eqphi} \\
\psi^{\prime\prime}+\frac{(d-1)+(d+1)\rho^2}{\rho(1+\rho^2)}\psi^\prime+\frac{(d-3)+(d-1)\rho^2}{\rho^2(1+\rho^2)}\psi&\nonumber\\ 
-(d-2)\frac{(1+q\rho\,\psi)(2+q\rho\,\psi)}{\rho^2(1+\rho^2)}\psi-q\frac{1+q\rho\,\psi}{\rho(1+\rho^2)}\varphi^2&=0\,. \label{eqpsi}
\end{align}
\end{subequations}
The large $\rho$ behavior of $\varphi$ and $\psi$ is
\begin{align}
\varphi=\frac{\varphi_{(1)}}{\rho^{d-\Delta}}+\frac{\varphi_{(2)}}{\rho^\Delta}+\dots, \\
\psi=\frac{\psi_{(1)}}{\rho}+\frac{\psi_{(2)}}{\rho^{d-1}}+\dots, \label{asymptotic psi}
\end{align}
where $\Delta>0$ is the largest of the two solutions to the equation $\Delta(\Delta-d)=m^2$.

\begin{figure*}[t]
\centering
\includegraphics[width=0.46\textwidth]{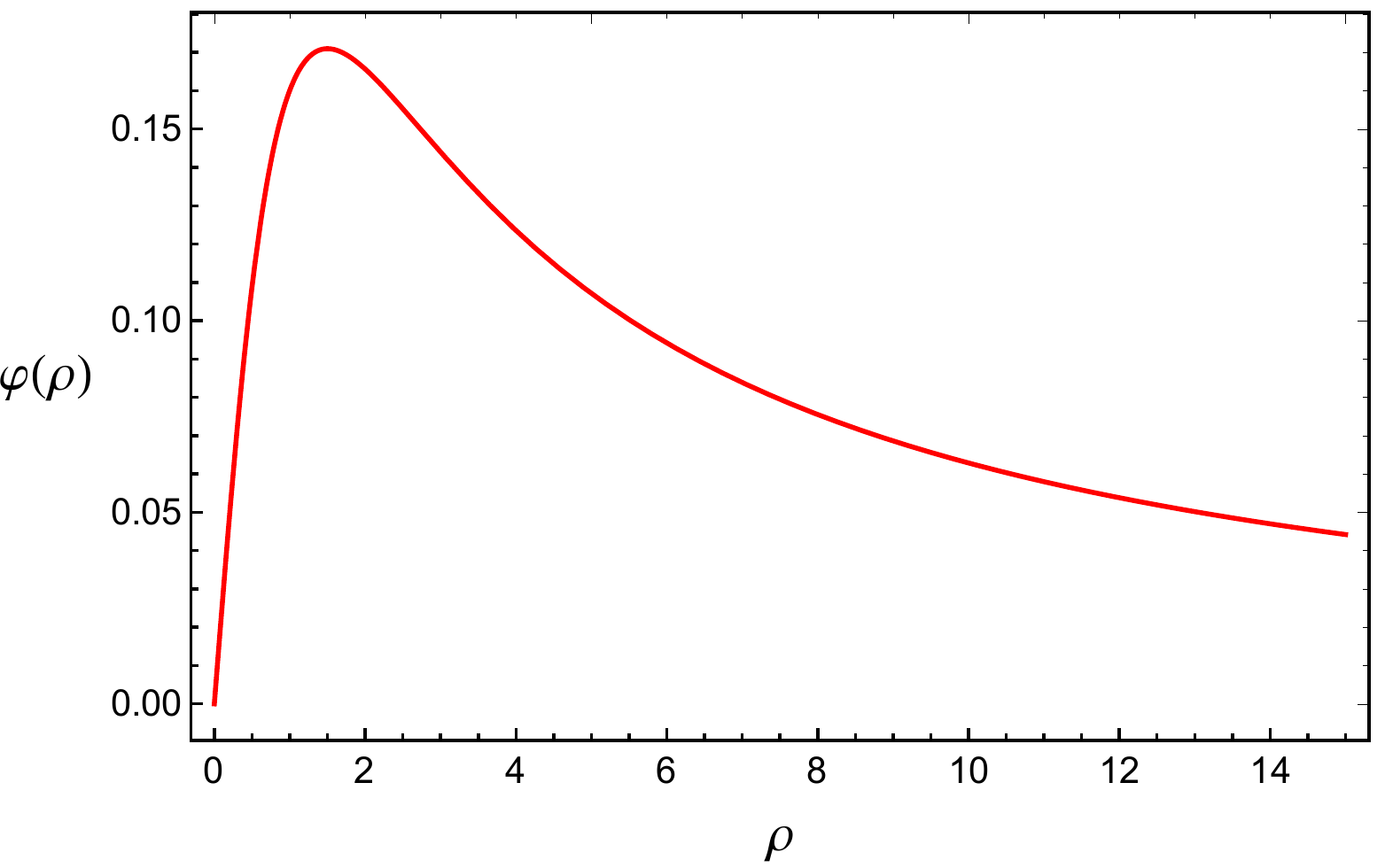} \hspace{0.5em}
\includegraphics[width=0.477\textwidth]{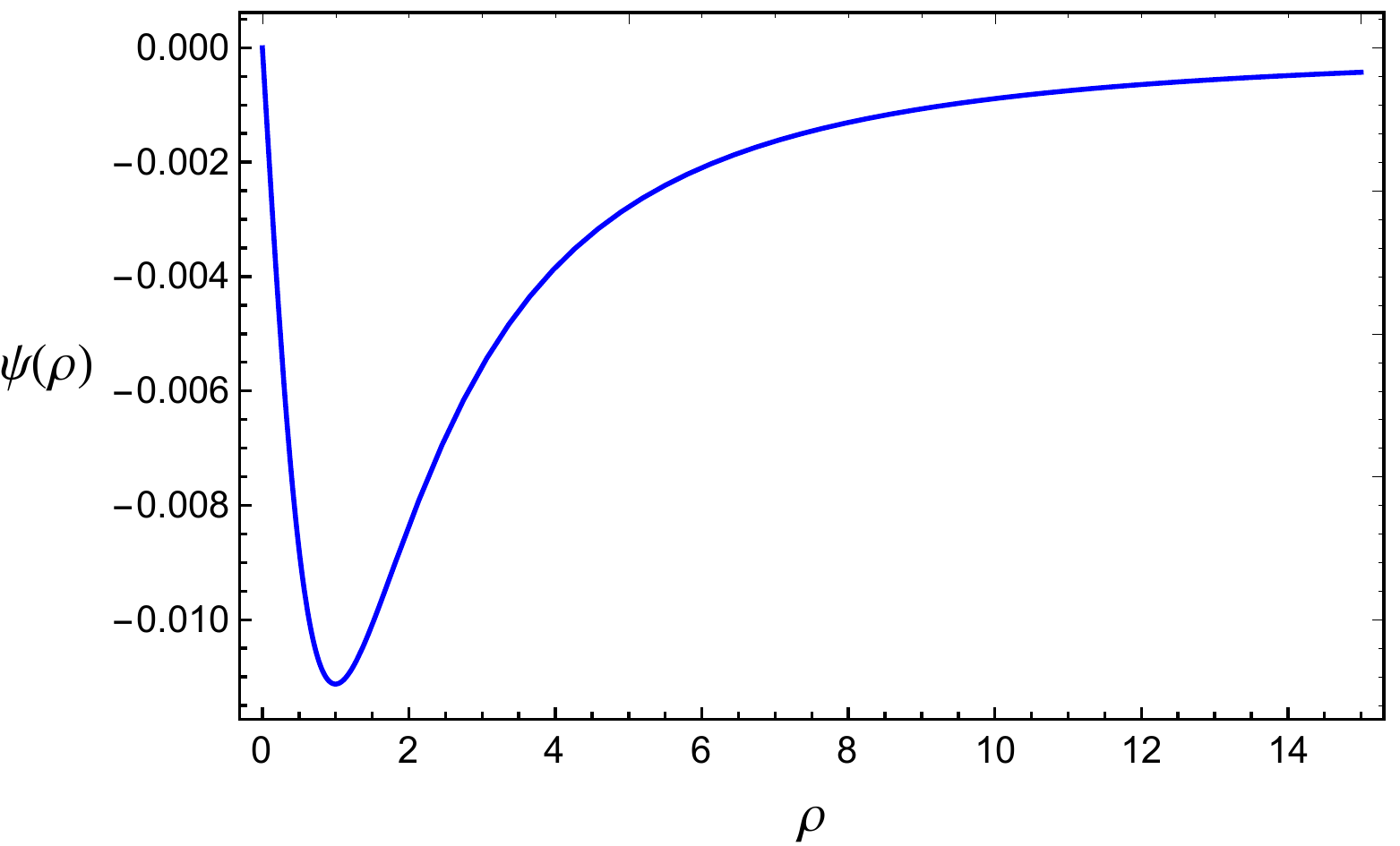}
\caption{Sample solutions for the scalar (left panel) and gauge field (right panel) profiles in global coordinates, for a potential $V(\varphi^2)=-\frac{d-1}{2}\varphi^2$, $q=1$ and boundary dimension $d=3$. Both solutions are regular everywhere. The gauge field satisfies the boundary condition $\psi_{(1)}=0$, while for the scalar field we imposed the mixed boundary condition $\varphi_{(2)}=f\varphi_{(1)}$ for $f=-1.8$.} \label{fig:global}
\end{figure*}

Since we are interested in the {\it spontaneous} (rather than explicit) breaking of the $SO(d)\times SO(d)$ symmetry, we must set to zero the sources of all the charged operators on the boundary. When it comes to the gauge field, this amounts to imposing $\psi_{(1)}=0$, given that the associated boundary currents are charged under the internal $SO(d)$ and their source is proportional to $\psi_{(1)}$. For the scalar field, instead, the expression for the boundary source in terms of the two falloffs depends on the form of $S_\text{bdy}$~\cite{Papadimitriou:2007sj}. In the absence of a temporal component of the gauge field (providing an effective negative mass), the only way for the scalar field to condense and spontaneously break the symmetry is through mixed boundary conditions\footnote{We are grateful to S.~Hartnoll for pointing this out.}~\cite{Witten:2001ua,Hertog:2004rz,Papadimitriou:2007sj}. This is reflected by the fact that the only consistent solution of Eqs.\ \eqref{eqphipsi} with $\psi_{(1)} = 0$ and either $\varphi_{(1)} = 0$ or $\varphi_{(2)} = 0$ is the trivial one, $\psi (\rho)  = \varphi (\rho) = 0$.

Mixed boundary conditions can be imposed provided the mass squared of the scalar field lies in the range $-d^2/4 < m^2 < 1-d^2/4$. In this case, the boundary term $S_\text{bdy}$ reads
%
\begin{align} \label{eq:Sbnd}
S_\text{bdy}&=-\frac{d-\Delta}{2}\int_{\rho\to\infty}\!\!\!\!\!\!\!\!\!d^dx\sqrt{-\gamma}\,|\vec\Phi|^2-\frac{2\Delta-d}{\nu+1}\,f\int d^dx\,|\vec\Phi_{(1)}|^{\nu+1} +{O}(\partial_\mu^2)\,,
\end{align}
where $f$ and $\nu$ are real dimensionless constants, $\gamma_{\mu\nu}$ is the induced boundary metric and ${O}(\partial_\mu^2)$ stands for terms that are at least quadratic in the boundary derivatives. The latter will be irrelevant for us since the background is independent of $x_\mu$ and we will perform a low energy expansion for the fluctuations. Note that the second term in \eqref{eq:Sbnd} is written directly in terms of the first falloff $\vec\Phi_{(1)} = \lim_{\rho \to \infty} \rho^{d-\Delta} \vec \Phi$ \cite{Faulkner:2010fh,Faulkner:2010gj}. This is because otherwise this term would not be finite on shell except in the conformal case $\nu=\Delta/(d-\Delta)$. The complete boundary term then renders the on-shell action finite and fixes the source for the operator $\vec {\cal O}(x)$ dual to the scalar field to be, at the background level, $J_\Phi\propto \varphi_{(2)}-f|\varphi_{(1)}|^{\nu-1}\varphi_{(1)}$. Our boundary conditions will then be $J_\Phi=0$.

The second boundary term corresponds to a multi-trace deformation of the boundary theory \cite{Witten:2001ua,Papadimitriou:2007sj}
\begin{equation} \label{multi trace}
f \cdot \int d^d x \sqrt{-g} \, \big| \vec {\cal O}(x) \big|^{\nu+1} \; ,
\end{equation}
with the parameter $f$ playing the role of a coupling constant.
This deformation should be analytic in the dual operator $\vec{\mathcal{O}}(x)$ for the conformal vacuum with $\langle\vec{\mathcal{O}}\rangle=0$ to exist, so that the CFT dual to our AdS model  can be regarded as a sensible UV completion of the EFT of solids. Thus the constant $\nu$ must be an odd positive integer. However, this restriction is not necessary as far as the spectrum of phonons is concerned, given that the latter arise as fluctuations in the spontaneously broken solid state, and hence away from the conformal vacuum. For this reason, in deriving the spectrum in the next section we will keep $\nu$ arbitrary, and actually consider in particular the value $\nu=\Delta/(d-\Delta)$, a choice that preserves conformal symmetry in the boundary theory, as a way to check our prediction for the dispersion relations of the phonons in a conformally invariant solid. For our numerical analysis of phase transitions in Section~\ref{sec:phase transition} we will make the choice $\nu=1$, corresponding to a relevant deformation (in $d=3$) which should therefore be present on naturalness grounds.

Since the equations of motion are of second order, their solutions depend a priori on four free parameters. Two of these are fixed by regularity at the center of AdS, and the other two by the boundary conditions we just discussed. Hence, Eqs.~\eqref{eqphipsi} admit a {\it unique} solution that satisfies our boundary conditions and is regular everywhere. This is consistent with the fact that for a solid on a sphere the background field configuration does not involve any free parameter---see discussion in Section~\ref{sec:solsphere}---the coupling constant $f$ being a fixed quantity in the microscopic theory rather than a parameter to be determined by boundary conditions. In Figure~\ref{fig:global}, we show this unique solution for a particular value of $f$ and with $d=3$, $\nu=1$.

This state of affairs should be contrasted with what happens for instance in the case of holographic superfluids~\cite{Hartnoll:2008vx,Hartnoll:2008kx,Herzog:2008he}. There, the background equations admit a one-parameter family of regular solutions consistent with the appropriate boundary conditions, because the chemical potential can be varied continuously. This is similar to what will happen for our solid in the flat limit, where the level of compression can be adjusted continuously and, correspondingly, the background equations will admit a one-parameter family of solutions.


\subsection{Gauge transformation} \label{sec:gauge transformation}

Before taking the flat limit, it will be convenient to act with a gauge transformation on our gauge and scalar field ansatz. As remarked earlier, the normal unit vector that appears in Eq.~\eqref{spherical ansatz 1} can be written as $\hat \rho=R\,\hat x_d$, with the rotation $R$ parametrized as follows:
\begin{align}
R=\prod_{i=1}^{d-1}\exp\left[\left(\theta_i-\frac{\pi}{2}\right)T^{i d}\right].
\end{align}
The factors of $\pi/2$ have been judiciously inserted in order to simplify the flat limit later on. If we now perform a gauge transformation with the group element $R^{-1}$ on the ansatz \eqref{spherical ansatz 1}, we obtain the physically equivalent field configuration
\begin{subequations} \label{new ansatz}
\begin{align}
\vec\Phi&=\varphi(\rho)\hat x_d, \\
A_M^{AB}&=\rho\,\psi(\rho)(R^{-1}\partial_M R\,\hat x_d)^T \,T^{AB}\,\hat x_d -\frac{1}{2q}\text{tr}\left(R^{-1}\partial_M R T^{AB}\right). \label{eq:Anew}
\end{align}
\end{subequations}
This transformation is analogous to the one that connects the 't Hooft-Polyakov monopole in the so-called hedgehog gauge to the Dirac monopole~\cite{Weinberg:2012pjx}.

With this new ansatz, the gauge field does not vanish anymore at the boundary because of the second term in Eq.\ \eqref{eq:Anew}. However, this  does {\it not} mean that we have turned on a source for the charged boundary currents associated with the gauge fields. Physics on the boundary cannot depend on any gauge transformation we perform in the bulk. In particular, the isometries on the sphere and the internal $SO(d)$ symmetries are still spontaneously broken. This will become obvious when we derive the low-energy spectrum of boundary excitations in the flat limit and show that the phonons are indeed gapless: if we had  broken the symmetries explicitly, these excitations would have been gapped.


\subsection{Flat limit}

We are now ready to implement the flat limit of the boundary theory. This requires a change from global coordinates to Poincar\'e patch in the bulk, which can be done by rescaling the coordinates as
\begin{align}
\rho=\mathcal{R} \,r\,,\qquad\tau=t/\mathcal{R}\,,\qquad\theta_i=\frac{\pi}{2}-\frac{x_i}{\mathcal{R}}\,,
\end{align}
and taking the large $\mathcal{R}$ limit~\cite{Bolognesi:2010nb}. By performing these replacements in Eqs.\ \eqref{eq:metric} and \eqref{new ansatz} we are taking the flat limit of our theory around the $x_d$-direction. The metric becomes now
\begin{align}
ds^2=-r^2dt^2+\frac{dr^2}{r^2}+r^2d\vec x^2 + O(\mathcal{R}^{-2})\,,
\end{align}
while the matter fields reduce to
\begin{align} 
\vec \Phi = \varphi(r)\hat x_d\,, \quad \;
A_M^{AB} = r\,\psi(r) T^{AB}_{id}\delta_M^i+O(\mathcal{R}^{-2})\,.  \label{eq:ansatz}
\end{align}
With a slight abuse of notation we have relabeled our profiles so that $\varphi(\mathcal{R}r)\to\varphi(r)$ and $r\psi(\mathcal{R}r) - 1 /(\mathcal{R}q) \to r\psi(r)$. Notice in particular that we have absorbed the constant $ 1 /(\mathcal{R}q)$ by shifting the first fall-off of $\psi (r)$, which is now non-zero and can be regarded as a free parameter in the same spirit as in Section~\ref{sec:solsphere}. In Appendix~\ref{app1} we present an alternative way of deriving the ansatz~\eqref{eq:ansatz}, without the need of passing through global coordinates.

The equations of motion in Poincar\'e patch are now considerably simpler and read
\begin{subequations}
\begin{gather}
\varphi^{\prime\prime}+\frac{d+1}{r}\varphi^\prime-(d-1)\frac{q^2\psi^2}{r^2}\varphi-2\frac{V^\prime(\varphi^2)}{r^2}\varphi=0, \label{eq:phi} \\
\psi^{\prime\prime}+\frac{d+1}{r}\psi^\prime+\frac{d-1}{r^2}\psi-\frac{q^2\varphi^2}{r^2}\psi-(d-2)\frac{q^2\psi^3}{r^2}=0. \label{eq:psi}
\end{gather}
\end{subequations}
Despite the fact that an analytic solution to these equations is not available, we will show in the next section that it is still possible to compute the action for the boundary Goldstones explicitly.
\\


\section{Phonons on the boundary}

We now apply the techniques developed in~\cite{deBoer:2015ija} and recently generalized in~\cite{Esposito:2016ria} to construct the boundary action for phonons. We will solve the equations of motion for the fluctuations of the matter fields around the background solutions discussed in the previous section, and show that the partially on-shell action correctly reproduces the sound speeds of phonons in a conformal solid.
\\

\subsection{Matter fields fluctuations} \label{sec:fluctuations}

We first introduce the fluctuations of the bulk fields. In particular, the standard approach~\cite{weinberg1995quantum} is to parametrize the gapless fluctuations as a {\it global} $SO(d)$ modulation of the symmetry breaking background with {\it local} transformation parameters. Thus, we write
\begin{align}
\vec\Phi=(\varphi +\sigma)\mathcal{M} \cdot \hat x_d,\quad A_M=\mathcal{M} \cdot (\bar A_M+\alpha_M) \cdot \mathcal{M}^{-1}\,,
\end{align}
with $\mathcal{M}=\exp(-\pi^i  T^{\,id})$. We also temporarily wrote the gauge field as a matrix in the adjoint representation, and indicated its background value as $\bar A_M$.

It is once again convenient to perform a gauge transformation and go to unitary gauge, where the fields are now given by
\begin{align}
\vec \Phi=(\varphi+\sigma)\hat x_d,\qquad A_M^{AB}=\bar A_M^{AB}+\delta A_M^{AB}\,.
\end{align}
With this gauge choice we have, at linear order in the fluctuations, $\delta A_M^{AB}=\alpha_M^{AB}-\frac{1}{q}\partial_M\pi^{AB}$, with $\pi^{id}=-\pi^{di}=\pi^i$ and $\pi^{ij}=\pi^{dd}=0$. At this point we still have the residual gauge freedom of applying a transformation generated by the $T^{ij}$ only, under which the scalar field is invariant. We use this to also fix the fluctuations so that $\delta A_r^{ij}=0$.

We are now ready to compute the action for the boundary Goldstones. The procedure will be similar to that outlined in~\cite{Esposito:2016ria}. We will first write down the linearized equations of motion for the fluctuations, at lowest order in boundary derivates. We will then solve all of them but the ones for the Goldstones fields, $\pi^i$. By plugging these solutions back into the quadratic action we will recover the action for the phonons of a conformal solid on the boundary.

The fluctuations $\pi^i$ always appear with a boundary derivative. Therefore, to implement the low energy limit of the theory we will consider the following scaling rules~\cite{deBoer:2015ija,Esposito:2016ria}
\begin{align}
\sigma,\,\delta A_\mu^{AB},\,\partial_r\sim O(1),\qquad\partial_\mu\sim O(\epsilon),\qquad\delta A_r^{id}\sim O(1/\epsilon)\,.
\end{align}
At lowest order in $\epsilon$ the equations of motion for the fluctuations read\footnote{As discussed in~\cite{Esposito:2016ria}, the rigorous way to take the low energy limit would be to first fix an IR cutoff, $r=\delta$, take the $\epsilon\to0$ limit and then send $\delta\to0$ at the end of the procedure. It was however found that the order of the limits was irrelevant.}
%
\begin{align}
\sigma^{\prime\prime}+\frac{d+1}{r}\sigma^\prime-(d-1)\frac{q^2\psi^2}{r^2}\sigma-2\frac{V^\prime+2\varphi^2 V^{\prime\prime}}{r^2}\sigma+2\frac{q^2\varphi\psi}{r^3}\delta A^{id}_i&=0, \\
\delta A^{id\,\prime\prime}_j +\frac{d-1}{r}\delta A^{id\,\prime}_j +\frac{q^2\psi^2}{r^2}\left[ \delta A^{jd}_i - (d-3)\delta A_j^{id} - 2\delta^i_j\delta A^{kd}_k \right]&\notag \\
-\frac{q^2\varphi^2}{r^2}\delta A^{id}_j+ 2\frac{q^2\varphi\psi}{r}\delta^i_j\sigma&=0, \\
\delta A_t^{id\,\prime\prime}+\frac{d-1}{r}\delta A^{id\,\prime}_t -(d-2)\frac{q^2\psi^2}{r^2}\delta A^{id}_t - \frac{q^2\varphi^2}{r^2}\delta A^{id}_t&=0, \label{eq:At} \\
\delta A_r^{id}&=0\,.
\end{align}
%
The very last equation tells us that the angular modes, $\pi^i$, are just Wilson lines of the gauge field fluctuations~$\alpha_r^{id}$,~{\it i.e.}
\begin{align} \label{wilson line}
\pi^i=q\int_0^r dr'\, \alpha_r^{id}(r')\,,
\end{align}
where we have chosen vanishing boundary conditions at the center of AdS. Similar results have already been found in several other contexts (see {\it e.g.}~\cite{Contino:2003ve,Sakai:2004cn,Nickel:2010pr,deBoer:2015ija,Esposito:2016ria}). The Goldstone bosons of the solid are to be identified with the boundary values of the bulk fields in Eq.~\eqref{wilson line}, {\it i.e.} $\pi_B^i(x_\mu)\equiv\pi^i(r=\infty,x_\mu)$.

It turns out that, in order to solve the remaining equations, we do not need the detailed knowledge of the background fields. We refer the reader to \cite{Esposito:2016ria} for details about the method. Let us start with Eq.~\eqref{eq:At}. It is easily shown that if we write $\delta A^{id}_t=c_t^i(x_\mu)r\psi(r)$, the latter reduces to the background equation~\eqref{eq:psi}, and thus it is automatically satisfied provided $\psi(r)$ is a valid solution. 
The standard procedure is now to impose vanishing Dirichlet boundary conditions for the gauge field fluctuations $\alpha_\mu^{id}$, both at $r=0$ and at $r=\infty$. We therefore arrive at
\begin{align} \label{eq:Atsol}
\delta A_t^{id}=-\frac{\partial_t\pi_B^i}{q\psi_{(1)}}\,r\,\psi(r)\,.
\end{align}
Given the prescribed boundary conditions, this will be the only regular solution~\cite{Witten:1998qj}.

To solve the other equations of motion it is convenient to decompose the fluctuations as follows:
\begin{align}
\delta A^{id}_j=\frac{1}{d-1}\delta^i_j\delta A^{kd}_k+\frac{1}{2}\mathcal{A}^i_j+\frac{1}{2}\mathcal{T}^i_j\,, \label{eq:param}
\end{align}
with $\mathcal{A}^i_j$ and $\mathcal{T}^i_j$ are respectively the antisymmetric and symmetric traceless combinations. When parametrized this way, the linearized equations of motion for the three terms in Eq.~\eqref{eq:param} decouple from each other.

Just like before one can show that the regular solution for the antisymmetric combination is proportional to the background profile, {\it i.e.} $\mathcal{A}^i_j=a^i_j(x_\mu)r\psi(r)$. Imposing double vanishing Dirichlet boundary conditions one gets
\begin{align} \label{eq:AA}
\mathcal{A}^i_j=-\frac{\partial_j\pi_B^i-\partial_i\pi_B^j}{q\psi_{(1)}}\,r\,\psi(r)\,.
\end{align}
Note that with our convention for the metric signature we do not need to distinguish between upper and lower spatial indices.

The equation for $\delta A^{id}_i$ is coupled to the radial fluctuation $\sigma$ and viceversa. One can show that they can be related to the derivatives of the background equations~\eqref{eq:phi} and \eqref{eq:psi}. In particular, the regular solutions turn out to be $\sigma=c(x_\mu)r\varphi^\prime(r)$ and $\delta A^{id}_i=-c(x_\mu)r^2\psi^\prime(r)/(d-1)$. Note that the behavior of $\phi(r)$ around $r = 0, \infty$ already ensures that $\sigma(r=0)=\sigma(r=\infty)=0$. Therefore, imposing the boundary conditions on $\delta A^{id}_i$ alone one gets
\begin{align} \label{eq:Atr}
\delta A^{id}_i&=\frac{\partial_i\pi_B^i}{q\psi_{(1)}}\,r^2\psi^\prime(r)\,, \\
\sigma&=-(d-1)\frac{\partial_i\pi_B^i}{q\psi_{(1)}}\,r\,\varphi^\prime(r)\,.
\end{align}

Let us finally analyze the traceless symmetric combination. Its equation of motion reads
\begin{align} \label{eq Tij}
\mathcal{T}^{i\prime\prime}_j+\frac{d-1}{r}\mathcal{T}^{i\prime}_j-\frac{q^2\varphi^2}{r^2}\mathcal{T}^i_j-(d-4)\frac{q^2\psi^2}{r^2}\mathcal{T}^i_j=0\,,
\end{align}
for which a solution in terms of background profiles is not available. However, even though we are unable to solve it for every $r$, it is still possible to constrain its near-boundary behavior. The general expression for large $r$ will be
\begin{align} \label{eq:AT}
\mathcal{T}^{ij}=-\frac{\partial^{\{i}\pi_B^{j\}}}{q}+\frac{\mathcal{T}_{(2)}^{ij}(x_\mu)}{r^{d-2}}+\dots\,,
\end{align}
where with $\{\,\cdots\}$ we represent the symmetric traceless part. 

Notice that the tensor structure of Eq.~\eqref{eq Tij} is trivial, in the sense that all components of $\mathcal{T}^i_j$ obey the same equation and do not mix with one another. Furthermore, since Eq.~\eqref{eq Tij} is an ODE in $r$ with $x$-independent coefficients,  all Fourier modes  $\tilde{\cal T}^i_j(r, k)$ also obey the same equation and do not mix with one another either. We can thus analyze the corresponding equation for a single function $f(r)$,
\begin{align}
f^{\prime\prime}+\frac{d-1}{r}f'-\frac{q^2\varphi^2}{r^2}f-(d-4)\frac{q^2\psi^2}{r^2}f=0\,,
\end{align}
and apply our considerations to each tensor component and each Fourier mode of $\mathcal{T}^i_j$ individually. This is a second-order ODE, and thus requires two boundary conditions to be integrated. For us these are the value of $f$ for $r \to \infty$, $f_{B}$, and the condition  of regularity at $r = 0$. With these boundary conditions the second falloff of $f$ for $r \to \infty$ is uniquely determined in terms of $f_B$, and, given the linearity of our ODE, must be proportional to it. Going back to ${\cal T}^i_j(r,x)$, we thus see that $\mathcal{T}_{(2)}^{ij}(x)$ must be proportional to $\partial^{\{i}\pi_B^{j\}}(x)$,
\begin{align} \label{eq:small_t}
\mathcal{T}_{(2)}^{ij}=-\frac{\psi_{(2)}}{q\psi_{(1)}}\,\lambda\,\partial^{\{i}\pi_B^{j\}}\,,
\end{align}
where $\lambda$ is an unknown integration constant, whose value should be determined by solving Eq.~\eqref{eq Tij}. The prefactor has been chosen for later convenience.


\subsection{Partially on-shell action}

Now that we have written the solution for the matter fields fluctuations (except for $\pi^i$), we can compute the partially on-shell action. Using the equations of motion the quadratic action reduces to the following boundary term:
\begin{align}
S^{(2)}&=-\int_{r\to\infty} \!\!\!\!\!\!\!\!\! d^dx\sqrt{-g}g^{rr}\bigg[ g^{\mu\nu}\frac{1}{2}\delta A_\mu^{id\prime}\delta A_\nu^{id}+ g^{\mu\nu}\frac{1}{4}\delta A_\mu^{ij\prime}\delta A_\nu^{ij} +\frac{1}{2}\sigma\sigma^\prime \bigg]+ S_\text{bdy}^{(2)}\,, \label{eq:line3}
\end{align}
where $S_\text{bdy}^{(2)}$ is the term of $S_\text{bdy}$ quadratic in the fluctuations. 

Given that $\pi^{ij}=0$ (see Section~\ref{sec:fluctuations}), the second term inside the brackets in Eq.~\eqref{eq:line3} does not contain the Goldstone fields, and can therefore be ignored for our purposes. The scalar fluctuation $\sigma$ enters in the third term in the brackets and in the boundary term $S_\text{bdy}^{(2)}$, which together give a contribution to the Goldstone action that is proportional to the dimensionless parameter
\begin{equation} \label{eq:constSigma}
\Sigma\equiv\frac{(d-1)^2(d-\Delta)^2(2\Delta-d)}{d-2}\left[\frac{\Delta}{d-\Delta}-\nu\right]\frac{f|\varphi_{(1)}|^{\nu+1}}{\psi_{(1)}\psi_{(2)}}\,,
\end{equation}
which vanishes precisely in the case of conformally invariant mixed boundary conditions, $\nu=\Delta/(d-\Delta)$. Using the solutions~\eqref{eq:Atsol}, \eqref{eq:AA}, \eqref{eq:Atr} and \eqref{eq:AT}, and collecting terms one obtains the quadratic action for the phonons of the dual solid,
\begin{subequations}
\begin{align}
S^{(2)}&=-\frac{d-2}{2}\frac{\psi_{(2)}}{q^2\psi_{(1)}}\int d^dx\bigg\{ \dot{\vec\pi}_B^2 - (\partial_k\pi_B^k)^2 -\frac{1}{2}\left[(\partial_j\pi_B^i)^2-(\partial_k\pi_B^k)^2\right]\notag \\
&\quad- \frac{\lambda}{2}\left[(\partial_j\pi_B^i)^2+\frac{d-3}{d-1}(\partial_k\pi_B^k)^2 \right] -\Sigma\,(\partial_k\pi_B^k)^2 \bigg\}\label{eq:Spi1}  \\
&=-\frac{d-2}{2}\frac{\psi_{(2)}}{q^2\psi_{(1)}}\int d^dx\left\{ \dot{\vec\pi}_L^2 +\dot{\vec\pi}_T^2 -\left[ 1+\lambda\frac{d-2}{d-1}+\Sigma \right] (\partial_j\pi_L^i)^2 - \frac{\lambda+1}{2}(\partial_j\pi_T^i)^2 \right\}\,, \label{eq:Spi2}
\end{align}
\end{subequations}
where in the second line we have again parametrized the Goldstone bosons in terms of transverse and longitudinal modes, $\vec\pi_B = \vec\pi_L + \vec\pi_T$---see Section~\ref{sec:EFT}. Numerical results show that $\psi_{(1)}$ and $\psi_{(2)}$ have opposite sign, therefore ensuring that, at least for some potentials, $\vec\pi_B$ has positive kinetic energy. From Eq.~\eqref{eq:Spi2} one reads off the two phonon speeds:
\begin{align} \label{eq:soundspeeds1}
c_T^2=\frac{\lambda+1}{2}\,,\qquad\text{ and }\qquad c_L^2=1+\lambda\,\frac{d-2}{d-1}+\Sigma\,.
\end{align}
%

Even though the precise values of the two phonon speeds are unknown at this stage, they do obey the relation
\begin{align} \label{eq:relc}
c_L^2=\frac{1}{d-1}+2\,\frac{d-2}{d-1}\,c_T^2\,,
\end{align}
%
whenever the parameter $\Sigma$ defined in \eqref{eq:constSigma} vanishes, which corresponds to the conformally invariant case. We have thus  reproduced the hallmark relation between longitudinal and transverse phonon speeds in a conformal solid, derived previously in Eq.~\eqref{conformal sound speeds}.

A comment is in order: The integration constant $\lambda$ is uniquely determined by solving the equation of motion $\eqref{eq Tij}$ and demanding that the solution be regular at $r=0$. Nevertheless, Eq.~\eqref{eq:relc} holds for {\it any} value of $\lambda$, in agreement with the conformal symmetry of the boundary theory. This is reassuring, since different scalar potentials in the bulk will likely yield different values for $\lambda$.

\section{Melting the solid} \label{sec:phase transition}

Now that we have argued that our $SO(d)$ magnetic monopole at zero temperature is indeed dual to a solid, it is natural to ask whether  it undergoes a phase transition when we turn on a finite temperature. In order to do that we first have to address some possible subtleties. First of all, we will move back to global coordinates (and hence to the description of a solid on a sphere), where we are confident that the global properties of our theory are all well defined. Second, we are now interested in hunting for a transition between a solution with a non-trivial scalar field and one with $\vec\Phi=0$. This means that we need to consider only multi-trace deformations \eqref{multi trace} that are analytic in $\vec {\cal O}$, that is, well defined also around the conformal vacuum. 
In light of the considerations made in Section~\ref{sec:setup}, this means 
considering mixed boundary conditions $\varphi_{(2)}=f\varphi_{(1)}^\nu$ with $\nu$ odd.

We reintroduce backreaction with the following metric ansatz
\begin{align}
ds^2=-(1+\rho^2)h(\rho)g(\rho)dt^2+\frac{h(\rho)}{g(\rho)}\frac{d\rho^2}{1+\rho^2}+\rho^2d\Omega_{d-1}^2\,.
\end{align}
From now on we will specialize to the $d=3$ case, and to a free tachyonic scalar field with mass $m^2=-2$, safely above the Breitenlohner-Freedman bound~\cite{Breitenlohner:1982bm}. We will also set $q=1$ for simplicity. The scalar and gauge field equations now read
%
\begin{subequations}
\begin{align}
\varphi^{\prime\prime}+\left(\frac{2+4\rho^2}{\rho(1+\rho^2)}+\frac{g^\prime}{g}\right)\varphi^\prime+\frac{2 h}{(1+\rho^2)g}\varphi-\frac{2h(1+\rho\psi)^2}{g\rho^2(1+\rho^2)}\varphi&=0\,, \\
\psi^{\prime\prime}+\left(\frac{2+4\rho^2}{\rho(1+\rho^2)}+\frac{g^\prime}{g}\right)\psi^\prime+\left(\frac{2}{1+\rho^2}+\frac{g^\prime}{\rho g}\right)\psi&\notag\\
-\frac{h(1+\rho\psi)(2+\rho\psi)}{g \rho^2(1+\rho^2)}\psi-\frac{h(1+\rho\psi)}{g\rho(1+\rho^2)}\varphi^2&=0\,.
\end{align}
\end{subequations}
%
If we turn off the backreaction by setting $h=g=1$, these equations reduce indeed to Eqs.~\eqref{eqphipsi} in the special case where $d = 3$ and $V' = -1$. Einstein's equations can instead be combined to obtain the following two first order equations for $g$ and $h$:
%
\begin{subequations}
\begin{gather}
g^\prime+\frac{1+3\rho^2}{\rho(1+\rho^2)}g-\frac{1+3\rho^2}{\rho(1+\rho^2)}h + \frac{\psi^2(2+\rho\psi)^2+2\varphi^2(1-\rho^2+2\rho\psi+\rho^2\psi^2)}{2\rho(1+\rho^2)}h=0\,, \\
h^\prime-\left(\frac{\rho}{2}\varphi^{\prime 2}+\frac{(\psi+\rho\psi^\prime)^2}{\rho}\right)h=0\,.
\end{gather}
\end{subequations}
%
Einstein's equations also include a second order one, which however can be derived from the ones above. Note that we have set the Planck mass $\mpl$ to unity; the fact that we can choose $\mpl=1=L$ by an appropriate rescaling of the fields and couplings is shown explicitly in Appendix \ref{app2}.

There are two possible ways to put our solid at finite temperature. One could either look for black hole solutions or simply compactify the time, {\it i.e.} study a thermal soliton. We will choose this second possibility, the reason being twofold. First of all, it can be shown numerically that for every given coupling $f$ there is always a temperature below which there are no black hole solutions with the required boundary conditions (see below). This is in contrast with the fact that solids at zero temperature do exist. Secondly, it is well known that the free energy of a black hole in $3+1$ dimensions scales as $F\propto N_c^2T^3$, where $N_c$ is the (large) number of colors of the boundary theory. On the other hand, when the fluctuations are turned off, the free energy of a solid should simply be constant (see {\it e.g.}~\cite{landau1980statistical}), which is indeed what happens to a thermal soliton for time-independent backgrounds.

We will now solve the full set of equations above. By requiring regularity of all the fields at the center of AdS one finds that there are only three free parameters, which can be taken to be $\varphi^\prime(0)$, $\psi^\prime(0)$ and $g(0)$. Close to the boundary, the asymptotic behavior of the fields is
\begin{subequations}
\begin{align}
\varphi&=\frac{\varphi_{(1)}}{\rho}+\frac{\varphi_{(2)}}{\rho^2}+\dots\,, \\
\psi&=\frac{\psi_{(1)}}{\rho}+\frac{\psi_{(2)}}{\rho^2}+\dots\,, \\
g &= g_{(0)}+\frac{g_{(0)}\varphi_{(1)}^2}{4\rho^2}+\frac{g_{(3)}}{\rho^3}+\dots\,, \\
h & = g_{(0)}-\frac{g_{(0)}\varphi_{(1)}^2}{4\rho^2}-\frac{2g_{(0)}\varphi_{(1)}\varphi_{(2)}}{3\rho^3}+\dots\,.
\end{align}
\end{subequations}
We will use the three free parameters to impose $g_{(0)}=1$, $\psi_{(1)}=0$ and $\varphi_{(2)}=f\varphi_{(1)}$. The last condition corresponds to deforming the boundary theory with a double trace operator, which in $d = 3$ is the only {\it relevant} deformation analytic in $\vec \Phi$.

Conveniently, our equations of motion exhibit the following symmetry
\begin{align}
t\to t/a,\qquad g\to ag,\qquad h\to a h,
\end{align}
which we can always employ to rescale any given solution so that it satisfies $g_{(0)}=1$. We are therefore left with just two initial conditions, $\varphi^\prime(0)$ and $\psi^\prime(0)$, which we use as shooting parameters to impose the last two boundary conditions on the matter fields. 

The free energy of our system can be obtained from the Euclidean on-shell action as $F=T S_E$, where the temperature is the inverse of the periodicity of time. Given that our background is time independent, this means that we can effectively set our system to any temperature. The complete on-shell action is
\begin{align}
S_E&=-\!\!\int d^4x_E\sqrt{g_E}\left[\frac{R}{2}+3+\mathcal{L}_m\right] \!+S_E^\text{GH}\!+\!S_E^\text{c.t.}\!+\!S_E^\text{bdy},
\end{align}
where $\mathcal{L}_m$ is the matter Lagrangian (see Eq.~\eqref{eq:Sm}), and $S_E^\text{bdy}$ is the Euclidean version of the action presented in Eq.~\eqref{eq:Sbnd}, with $\nu=1$. Moreover $S_E^{GH}$ and $S_E^\text{c.t.}$ are the Gibbons-Hawking~\cite{Gibbons:1976ue} and counterterm~\cite{Bianchi:2001kw} actions, necessary to fix the variational problem as well as the UV divergences of the on-shell gravity action---see also~\cite{natsuume2015ads} for a recent textbook treatment. In our case they are given by
\begin{align}
S_E^\text{GH}\!&=-\!\int_{\rho\to\infty}\!\!\!\!\!\!\!\!\!d^3x_E\,\sqrt{\gamma_E} \, K, \\
S_E^\text{c.t.}\!&=\!\frac{1}{2}\int_{\rho\to\infty} \!\!\!\!\!\!\!\!\!d^3x_E\sqrt{\gamma_E}\Big[ 4+\mathscr{R} + \mathscr{R}^{\mu\nu}\mathscr{R}_{\mu\nu} - \frac{3}{8}\mathscr{R}^2 \Big].
\end{align}
Here, $K = n^\rho\partial_\rho \log \sqrt{\gamma_E}$ is the extrinsic curvature of the boundary, $\gamma_E$  the Euclidean induced  metric, $n^\rho=1/\sqrt{g_{\rho\rho}}$ the $\rho$-component of the vector normal to the boundary, and $\mathscr{R}_{\mu\nu}$ the Ricci tensor built out of the induced metric. 
The order parameter for our solid state is given by the vev of the operator dual to the scalar field $\vec\Phi$. In order to look for possible phase transitions we then need to compare our system to a black hole solution with $\vec\Phi=0$. In the absence of a source for the currents dual to the gauge field (see Sec. \ref{sec:gauge transformation}), the only possible solution with no scalar field is the one where the gauge field is also zero. We then consider the simple Schwarzschild-AdS (SAdS) black hole. It is well know that such a solution is dual to a fluid on the boundary theory~\cite{Bhattacharyya:2008jc,Nickel:2010pr,deBoer:2015ija,Crossley:2015tka}.

\begin{figure}[t]
\centering
\includegraphics[width=0.6\textwidth]{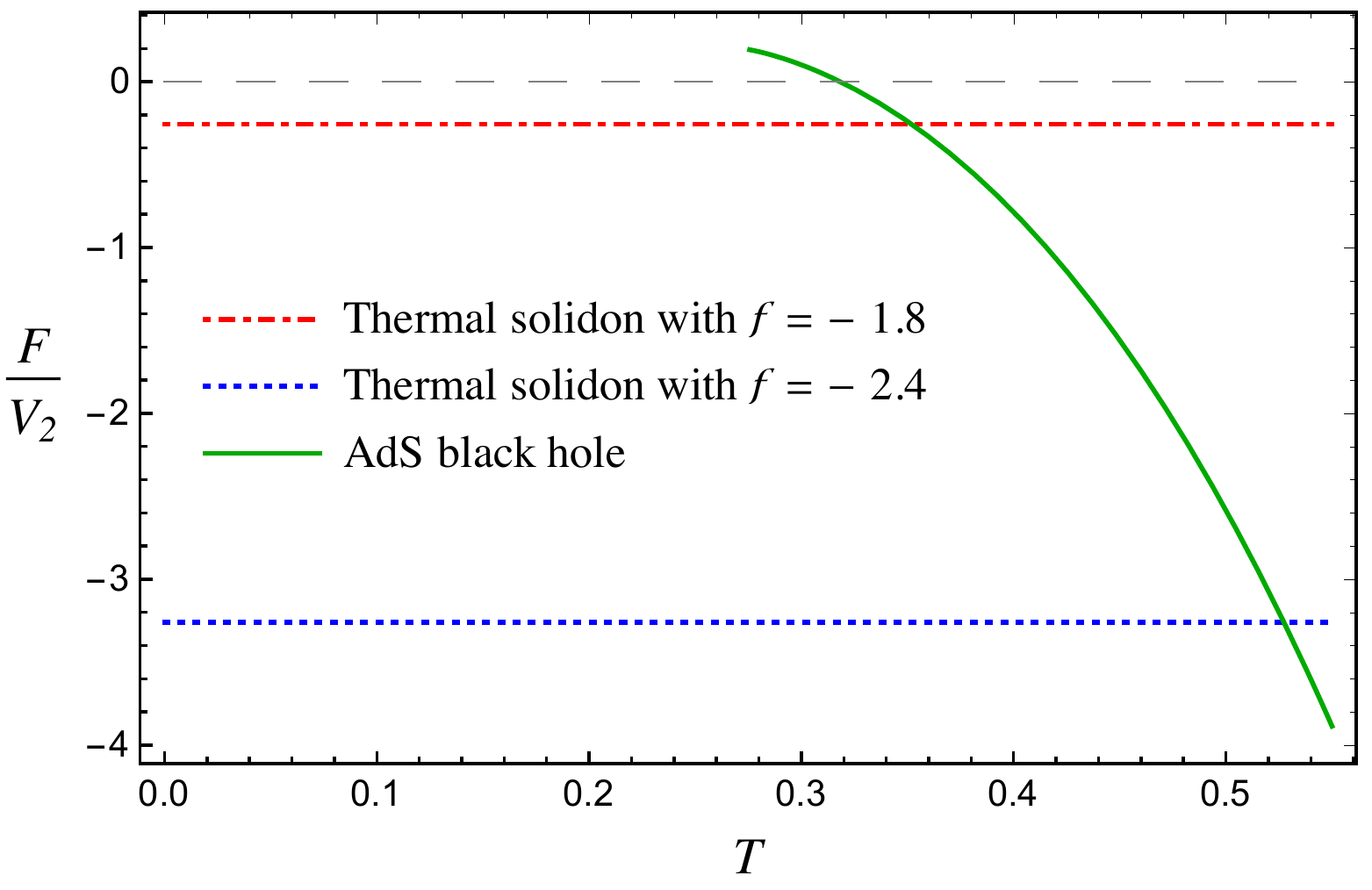}
\caption{Comparison between the free energies of the thermal solidon with $f=-1.8$ (red, dot dashed), and with $f=-2.4$ (blue, dotted) and of a large AdS black hole (green, solid). Below $T=\sqrt{3}/2\pi$ the black hole solution ceases to exist. It is clear that for low enough temperatures our solid configuration is preferred.} \label{fig:F}
\end{figure}

In Figure~\ref{fig:F} we show the free energies of our thermal solidon for two different values of the coupling $f$, compared with that of a large SAdS black hole. It is clear that at low enough temperature our solid configuration is favored with respect to the black hole one. Moreover, the derivative of the free energy is discontinuous at the critical temperature, which is the hallmark of a first order phase transition. The dual field theory interpretation of this is indeed that of a solid-to-liquid melting.

Before closing this section let us comment on the role of phonons. At low temperatures, the gas of phonons will contribute to the free energy of a solid with a term $F_\pi\propto 2 \times T^3$, with $2$ being the number of independent phonon polarizations in $d = 2+1$ dimensions~\cite{landau1980statistical}. This contribution does not appear in our free energy since we are setting all the fluctuations to zero, but a more detailed analysis should include it. However, provided that the sound speeds are not too low, the large $N_c^2$ coefficient of the black hole free energy should still ensure the existence of a first order phase transition.

Another important question is whether the parameters of the solutions we have studied are such that the dual systems can indeed be regarded as solids in the sense explained in Section \ref{sec:conformal solids}. This requires a parametric separation between the IR and UV cutoffs of the EFT, allowing in principle for the existence of Goldstone modes related to the solid's symmetry breaking pattern. The relevant IR scale is the inverse radius of the sphere, $1/{\mathcal{R}}$, while the UV scale is set by the energy density $\langle T_{tt}\rangle$ of the boundary theory divided by an appropriate factor of $N_c$ (due to the fact that $1/N_c$ plays the role of a coupling constant in large $N_c$ theories, thus modifying the na\"ive UV cutoff that one would infer directly from the EFT). For $d=3$ the requirement that the IR and UV scales are widely separated amounts therefore to $\langle |T_{tt}| \rangle /N_c^2 \gg 1/{\mathcal{R}}^3$. We have checked that the above solutions in fact do not satisfy this, but it turns out that this is just a limitation of our numerics, as we argue in Appendix \ref{app3} that there exist parameters for which the system indeed meets this requirement.

\section{Conclusions}

The goal of this paper was to  construct a gravitational dual of a strongly coupled theory in a solid state. Our bulk theory consists of Einstein gravity coupled to a $SO(d)$ Yang--Mills theory and a real scalar field in the fundamental representation. We showed that such a theory admits a monopole solution---the {\it solidon}---in the presence of mixed boundary conditions for the scalar field. The bulk scalar is associated with a boundary operator whose expectation value realizes the spontaneous symmetry breaking pattern of a solid on a sphere. Our main argument in support of this claim was the derivation of the correct phonon spectrum of the boundary theory, which we computed in the flat, probe limit. We showed that the effective low-energy theory on the boundary contains $d-1$ gapless modes $\pi^i_B$---the phonons of the solid---and is, at least for some potentials, ghost-free. Moreover, when the mixed boundary conditions preserve conformal invariance, the longitudinal and transverse phonon speeds satisfy the universal relation \eqref{conformal sound speeds}, characteristic of a conformal solid. 

We also studied the possible phase transitions of our solidon at finite temperature. To this end, we reintroduced the backreaction and compared the solidon free energy with that of a SAdS black hole, which is known to describe a fluid state on the boundary. We found that a first order phase transition takes place between the solidon at low temperatures and SAdS at high temperatures: this is dual to the melting of our solid.

It is also interesting to note how our analysis shows that, when the temperature is lowered, the SAdS-to-solidon transition happens {\it before} the Hawking-Page one. This suggests that there might exist strongly coupled theories that at low temperatures undergo a transition to a solid state rather than to a confined one.


The generality of the solid dual to the bulk theory we presented is yet to be understood. The integration constant $\lambda$, and consequently the phonon speeds, will most likely depend on the falloffs of the background profiles. We therefore expect to obtain different phonon speeds for different scalar potentials. A complete analysis requires introducing gravitational backreaction and studying (numerically) the poles of the scalar two-point function. From this viewpoint, it may also be interesting to study generalizations of our bulk theory along the lines of what was done in Ref.~\cite{Franco:2009yz} for holographic superconductors.

At this stage, it is hard to foresee what practical applications our construction may give rise to. Conformal solids necessarily have an energy density that is comparable with their pressure, and are therefore relativistic. Systems made in the lab are on the other hand usually non-relativistic---they are made of atoms with binding and interaction energies that are negligible compared to their atomic mass~\cite{Nicolis:2015sra}, and as a result their sound speed is usually much smaller than the speed of light. Nevertheless, when conformal symmetry is broken by a relevant deformation, there might be an holographic dual  for which the sound speeds of our solid are both non-relativistic. In this case, it would be of great interest to study, for example, the transport properties of the boundary theory, in order to determine whether or not it could describe any experimental system. Moreover, given that one of the main motivations of the ``AdS/CMT'' program is to gain insight into the phase structure of strongly coupled systems, having  concrete gravity duals of  solids is certainly a desirable addition to the picture.

It is not obvious what the relation is between
the bulk theory we have considered here and the models of holographic massive gravity~\cite{Vegh:2013sk,Blake:2013bqa} that have been studied as duals of solids~\cite{Alberte:2015isw,Alberte:2016xja, Alberte:2017cch}. There, the emphasis is on the short-distance breaking of translations on the boundary theory, which is interpreted in the bulk as a breaking of general covariance. Our starting point has been instead the symmetry breaking pattern of the low-energy EFT of solids, which involves certain internal symmetries that are needed for the low-energy dynamics to be correctly reproduced, but whose physical origin is still quite mysterious \cite{Nicolis:2015sra}. For the time being, the two approaches seem to be quite different from each other, but it would be interesting to understand whether any relation exists between them.


We remark that holographic models with spontaneous breaking of spatial translations have also been previously studied for instance in~\cite{Bolognesi:2010nb,Sutcliffe:2011sr}, where magnetic monopoles were considered as duals of systems with crystalline structure. A related analysis of crystalline solids within our framework would be very natural given that the EFT of solids can be systematically generalized to include anisotropies~\cite{Kang:2015uha}. 

Lastly, it would be interesting to numerically study the phase space of our holographic solids, by varying the different couplings and free parameters of the theory, both on the flat patch and on the sphere. We hope to address these questions in future work.


\begin{acknowledgments}

We are grateful to L.~Alberte, C.~Asplund, M.~Baggioli, G.~Chiriac\'{o}, L.~Delacr\'etaz, F.~Denef, S.~Hartnoll, G.~Horowitz, L.~Hui, K.~Jensen,  E.~Mitsou, R.~Monten, I.~Papadimitriou, A.~Pilloni, R.~Rattazzi, and C.~Toldo for interesting and helpful discussions. We would also like to acknowledge R.~Krichevsky for collaboration in the early stages of this project. A.N.~is particularly grateful to R.~Sundrum for illuminating discussions. A.E.~is grateful to K.~Wenz for useful suggestions on computational aspects. 

This work has been supported by the US Department of Energy under contracts DE-FG02-11ER41743, DE-FG02-92-ER40699, DE-SC0013528, and DE-SC0011941, and by the European Research Council under the European Community's Seventh Framework Programme (FP7/2007-2013 Grant Agreement no.\ 307934, NIRG project).  R.P. acknowledges the hospitality of the Sitka Sound Science Center during the final stages of this work.

\end{acknowledgments}

\appendix


\section{An alternative derivation of the bulk fields ansatzes} \label{app1}

We now show how the symmetry breaking ansatz~\eqref{eq:ansatz} can also be derived without invoking an underlying spherical manifold for the solid. This will require us to formulate our EFT of solids  in a slightly  different way. The equivalent results obtained with the two approaches can be regarded as a consequence of the approximate Poincar\'{e} symmetry shared by a sphere near its poles.

It is well known that the algebra for the Euclidean group $ISO(d-1)$ can be obtained from a contraction~\cite{inonu1953contraction} of the $SO(d)$ algebra.  In particular, if we separate the generators in those that transform as a vector, $T^{id}$, and those that transform as a tensor, $T^{ij}$, their commutation relations read
\begin{align}
\big[T^{ij},T^{km}\big]&=\delta^{ik}T^{jm}+\delta^{jm}T^{ik} -(k \leftrightarrow m),\\
\big[T^{ij},T^{kd}\big]&=\delta^{ik}T^{jd}-\delta^{jk}T^{id}, \\
\big[T^{id},T^{jd}\big]&=T^{ij}.
\end{align}
If we now define $T^{id}=\zeta P^i$ and take the $\zeta\to\infty$ limit, the $SO(d)$ algebra reduces to the $ISO(d-1)$ one, with $P^i$ the momentum generators. Note that while in the previous sections $\mathcal{R}$ was the radius of the spherical spacetime manifold, here we are always in flat space and $\zeta$ has no geometrical interpretation.

In taking this limit one also has to make sure to have a multiplet that correctly transforms as a solid. Let us consider a $d$-dimensional multiplet $\vec\psi$ in the fundamental representation of $SO(d)$. Under an infinitesimal transformation with parameters $\theta^{AB}$ generated by the rescaled operators $T^{ij}$ and $P^i$, the variations of the multiplet are
\begin{align}
\delta \psi^i=\frac{1}{2}\theta^{ij}\psi_j+\frac{1}{\zeta}\theta^{id}\psi_d,\quad\text{ and }\quad\delta\psi_d=\frac{1}{\zeta}\theta^{id}\psi_i.
\end{align}
If now we assume that the $d$-component of the multiplet takes a vev $\langle\psi_d\rangle=\zeta$, one immediately sees that when $\zeta\to\infty$, $\psi_d$ remains invariant while $\psi_i$ transforms under $ISO(d-1)$, hence realizing the symmetry breaking pattern of a solid. We can therefore identify $\psi_i=\phi_i$ (see Section~\ref{sec:EFT}). It is interesting to note that the fact that translations are non-linearly realized in $ISO(d-1)$ is in this language seen as the consequence of the spontaneous breaking of $SO(d)$ due to the large vev acquired by $\psi_d$.

We can now parametrize the complete field in terms of fluctuations around this background, {\it i.e.} by writing
\begin{align} \label{vecpsi}
\vec\psi(x)=\mathcal{O}(x)\vec\Phi=\left(\phi_1(x),\dots,\phi_{d-1}(x),\zeta\right)^T,
\end{align}
with
\begin{align}
\vec \Phi=\zeta\hat x_d,\qquad\mathcal{O}(x)=\exp\left(\phi_i(x)P^i\right).
\end{align}
Note that Eq.~\eqref{vecpsi} is true in the $\zeta\to\infty$ limit. It is the flat space analogue of Eq.~\eqref{EFT vev sphere}.

At this point, one would typically write the solid lagrangian in terms of the invariants built out of the matrix $B^{ij}=\partial_\mu\phi^i\partial^\mu\phi^j$. However, we note that
\begin{align}
\partial_\mu\vec\psi=\mathcal{O}\left( \mathcal{O}^{-1}\partial_\mu\mathcal{O}\vec \Phi \right)\equiv\mathcal{O} D_\mu\vec \Phi,
\end{align}
where in the first equality we have used the fact that $\vec\Phi$ is a constant field.
The same theory can then be written in terms of the invariants built out of $\tilde B^{ij}=(D_\mu \Phi)^i (D^\mu \Phi)^j$. However, on the background we have
\begin{align}
\langle D_\mu \vec \Phi\rangle=\langle \mathcal{O}^{-1}\partial_\mu\mathcal{O}\vec \Phi\rangle=\alpha \delta_\mu^i P^i\vec \Phi=\alpha \delta_\mu^i T^{id}\hat x_d.
\end{align}
Which means that theory for our solid can also be expressed in terms of a single scalar field with a vacuum expectation value $\langle\vec\Phi\rangle\propto\hat x_d$, and a constant gauge field $A_i^{AB}=\alpha\, T^{AB}_{id}$. This is indeed the ansatz found in Eq.~\eqref{eq:ansatz} in the Poincar\'e patch limit.

\section{Setting $\mpl = L \equiv 1$} \label{app2}

In this appendix, we will show how to effectively set $\mpl = L = 1$ by an appropriate rescaling of fields and couplings. Our starting point will be the $(3+1)$D action with all powers of $\mpl$ and $L$ restored, namely

%
\begin{align} \label{S tot app B}
S&=\int d^4x\sqrt{-g}\bigg[ \frac{\mpl^2}{2} R +\frac{3\mpl^2}{L^2} - \frac{1}{2}\,D_M\vec\Phi\cdot D^M\vec\Phi - V\big(|\Phi|^2\big) - \frac{1}{8}\,F_{MN}^{AB}F^{AB\,MN}\bigg] \notag \\
&\quad+ S_\text{GH+c.t.+bdy}\,,
\end{align}
with $D_M = \partial_M - i q A_M$, $F_{MN} = \partial_M A_N - \partial_N A_M - i q [A_M, A_N]$, 
\begin{align}
S_\text{GH+c.t.+bdy} &= \int_{\rho\to\infty}\!\!\!\!\!\!\!\!\! d^3x \sqrt{- \gamma} \bigg[ \mpl^2 K  -\frac{\mpl^2}{2L} \Big[ 4 +L^2\mathscr{R} + L^4\Big(\mathscr{R}^{\mu\nu}\mathscr{R}_{\mu\nu} - \frac{3\mathscr{R}^2}{8}\Big) \Big] \notag \\
&\quad-\frac{3-\Delta}{2 L}\,|\vec\Phi|^2  -\frac{2\Delta-3}{\nu+1}\,\frac{f\,|\vec\Phi_{(1)}|^{\nu+1}}{\sqrt{- \gamma}} \bigg] \,,
\end{align}
and with $K$ and $\mathscr{R}$ the extrinsic and intrinsic curvatures of the boundary respectively.
%
When we neglect backreaction, the background AdS metric reads
\begin{equation}
	ds^2 = -\left(1+\frac{\rho^2}{L^2} \right)d\tau^2+\frac{d\rho^2}{1+\frac{\rho^2}{L^2} } + \rho^2 d\Omega^2.
\end{equation}

Let us now introduce the rescaled fields
\begin{subequations}
\begin{align}
 	\Phi^A &\equiv \mpl \tilde \Phi^A \\
 	A_M &\equiv \mpl L \, \tilde  A_M \\
 	g_{MN} & \equiv L^2 \tilde g_{MN} ,
\end{align}
\end{subequations}
the rescaled couplings
\begin{subequations}
\begin{align}
 	q &\equiv \tilde q / \mpl L \\
 	f &\equiv \tilde f \mpl^{1- \nu } L^{2-(3-\Delta)(\nu+1)},
\end{align}
\end{subequations}
and the rescaled scalar potential 
\begin{equation}
	V\big(|\Phi|^2\big) \equiv  \tilde V (|\tilde{\Phi}|^2) \, \mpl^2 /L^2 .
\end{equation}
For instance, in the case of a quadratic potential $V = \tfrac{1}{2} m^2 |\vec\Phi|^2$, our rescaled potential would simply be $\tilde V = \tfrac{1}{2} \Delta (\Delta - 3) |\tilde \Phi|^2$, with the $\Delta$ the scaling dimension of the boundary operator associated with $\vec \Phi$. 

It is also convenient to perform a diffeomorphism and introduce new radial and time variables defined by $\tilde \rho \equiv \rho / L$, $\tilde \tau \equiv \tau/ L$. In these new coordinates,  the components of $\tilde g$ in the absence of backreaction are simply
\begin{equation} \label{rescaled metric}
	\tilde g_{MN} = \text{diag} \left[ - \left( 1+\tilde \rho^2\right), \frac{1}{1+\tilde \rho^2}, \, \tilde \rho^2, \, \tilde \rho^2 \sin \theta \right] .
\end{equation} 

Then, it is easy to check that these rescalings allow us to rewrite the action \eqref{S tot app B} as
\begin{equation} \label{eq:Stilde}
	S[\Phi, A, g] = \mpl^2 L^2 \tilde S [\tilde \Phi, \tilde A, \tilde g] ,
\end{equation}
where $\tilde S$ is an action of the same form as $S$, but with $\mpl \to 1, L \to 1, q \to \tilde q$ and $f \to \tilde f$. To prove this, it is also helpful to remember that $\vec \Phi_{(1)} = \lim_{\rho \to \infty} \rho^{3-\Delta} \vec \Phi$.

This shows that, rather than minimizing the action $S$, one can always choose to minimize the action $\tilde S$ since the two are proportional to each other. Both $L$ and $\mpl$ do not appear in $\tilde S$ nor in the background metric \eqref{rescaled metric}. There is however still a remnant of $L$ in the mixed boundary conditions for the scalar field. To illustrate this, consider for simplicity the special case $\Delta = 2$, $\nu =1$ that is relevant for the phase transition studied in Sec.\ \ref{sec:phase transition}. In this case, $\tilde f = f$ and it has dimensions of $[\text{length}]^{-3}$. It is then easy to show that imposing the mixed boundary condition $\varphi_{(2)} = f \varphi_{(1)}$ on the scalar profile $\varphi(\rho)$ defined by $\vec \Phi = \varphi(\rho) \hat \rho$ is tantamount to requiring that the falloffs of $\tilde \varphi(\tilde \rho)$ satisfy the condition $\tilde \varphi_{(2)} = (\tilde f L^3) \tilde \varphi_{(1)}$. Thus, it is still helpful to work in units such that $L=1$, which is why we have made this choice throughout the paper. Moreover, in setting $\mpl =1$ we have effectively worked with $\tilde \Phi^A, \tilde A_M, ...$ rather than $\Phi^A, A_M, ...$ all along, but have omitted the ``tildes'' to avoid overburdening the notation.

\section{The UV cutoff of the boundary theory} \label{app3}

In this appendix we compute the energy density of the boundary theory and, consequently, its UV cutoff. As discussed in Section~\ref{sec:conformal solids}, our definition of solid only requires the existence of long distance/low energy modes obeying the necessary symmetries. When working on a sphere, there is a natural IR cutoff given by the inverse of its radius, and it is therefore important to have a large separation between the IR and UV cutoffs, in order to have a regime where these degrees of freedom exist.

Here we show that the numerical results reported in Section~\ref{sec:phase transition} are actually obtained in a context where this separation is not large enough. Nevertheless, we also argue that the results we found can be extrapolated to the case where the UV cutoff is much larger that the IR one, and hence our conclusions are still valid.
\begin{figure}[t!]
\centering
\includegraphics[width=0.6\textwidth]{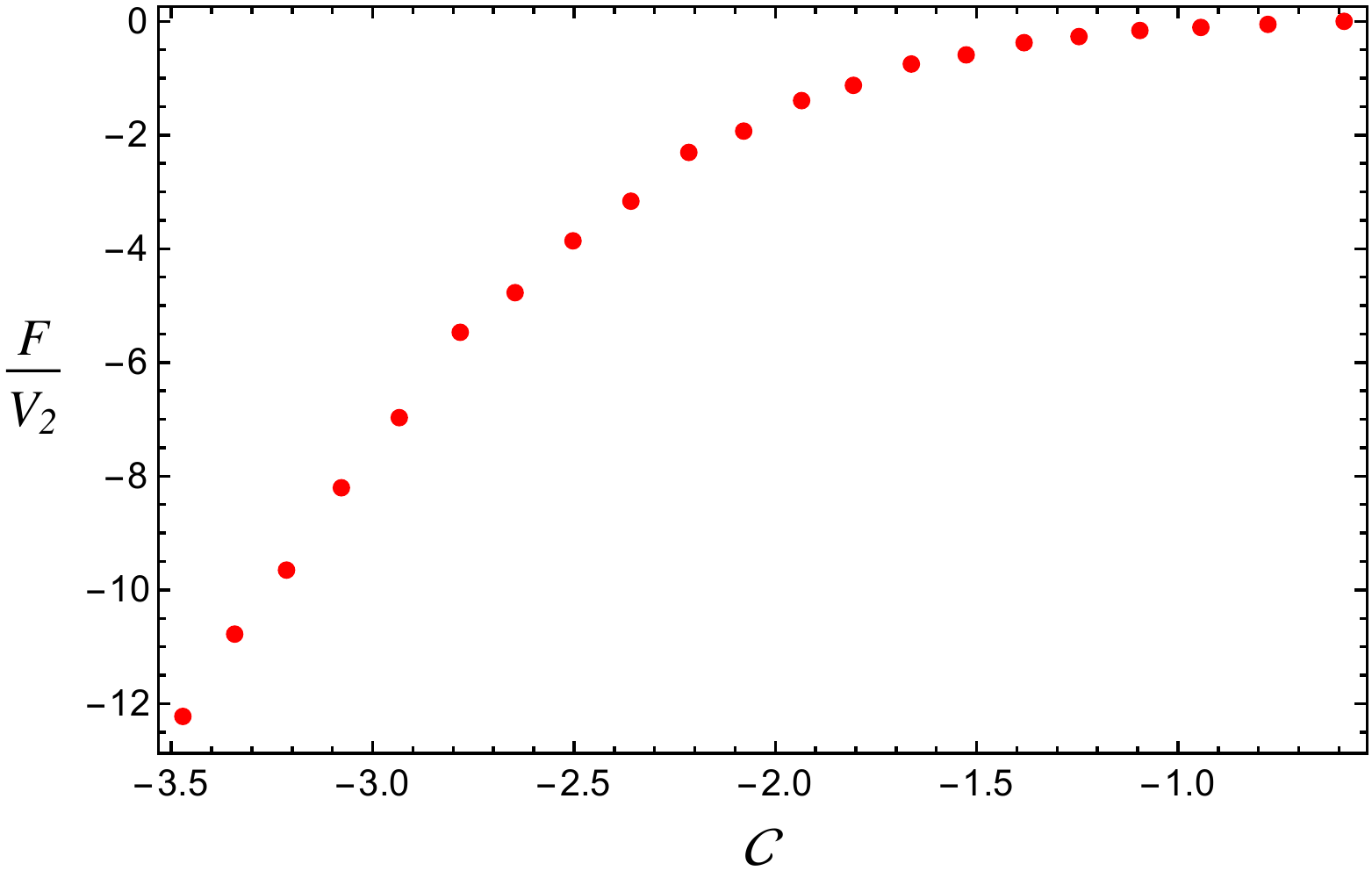}
\caption{Dependence of the free energy on the constant $\mathcal{C}$ defined in \eqref{eq:Cdef}. It is clear that when the magnitude of the latter one becomes larger the free energy become more negative. One should keep in mind that, for a given coupling $f$, the solution to the equations of motion is unique---see Sections~\ref{sec:setup} and \ref{sec:phase transition}. Different values of $\mathcal{C}$ therefore do not corresponds to different states of the same solid but rather to solids obtained from different theories.} \label{fig:FvsC}
\end{figure}

First, we find the energy density of our solid when its underlying manifold is a sphere of radius $\mathcal{R}$. In order to do that, it is convenient to write our metric in the Graham-Fefferman gauge~\cite{deHaro:2000vlm}, where it takes the form
\begin{align}
ds^2=\frac{L^2}{u^2}\left(du^2+H_{\mu\nu} dx^\mu dx^\nu\right).
\end{align}
One can show that, close to the boundary, the right change of coordinates is
\begin{align}
\rho&=\frac{L}{u}\Bigg( 1-\frac{2+\varphi_{(1)}^2}{8}\frac{u^2}{L^2} -\frac{3 g_{(3)}+2 g_{(0)}\varphi_{(1)}\varphi_{(2)}}{18 g_{(0)}}\frac{u^3}{L^3}+O(u^4/L^4) \Bigg).
\end{align}
In $d=2+1$ dimensions, the stress-energy tensor of the boundary theory is related to the third falloff of $H_{\mu\nu}$ by~\cite{deHaro:2000vlm}
\begin{align}
\langle T_{\mu\nu}\rangle = \frac{3}{2}\mpl^2L^2 H_{\mu\nu}^{(3)}.
\end{align}
In the case of a sphere of radius $\mathcal{R}$, we read off the falloffs by extracting a conformal factor $(\rho/\mathcal{R})^3$. In this case, the energy density of our solid turns out to be
\begin{align} \label{eq:Ttt}
\langle T_{tt} \rangle =\frac{3}{2} \mpl^2L^2\frac{\mathcal{C}^3}{\mathcal{R}^3}\sim N_c^2\frac{\mathcal{C}^3}{\mathcal{R}^3},
\end{align}
where $\mathcal{C}$ is given by
\begin{align} \label{eq:Cdef}
\mathcal{C}\equiv-\left(\frac{2}{3}\frac{ g_{(3)}}{ g_{(0)}}-\frac{8}{9}\varphi_{(1)}\varphi_{(2)}\right)^{1/3}.
\end{align}
This quantity is dimensionless and built only out of the falloffs obtained from the action $\tilde S$ in Eq.~\eqref{eq:Stilde}, which is exactly what we can compute numerically.

One could na\"ively think that, in order to have a large separation between the UV and IR cutoffs, it should be $\langle |T_{tt}| \rangle \gg 1/\mathcal{R}^3$, which would always be satisfied thanks to the large $N_c^2$ factor in Eq.~\eqref{eq:Ttt}. However, one has to keep in mind that $1/N_c^2$ plays the role of a coupling constant of the boundary theory, and therefore one really needs $|\mathcal{C}| \gg1$.

It is possible to check that the two free energies reported in Figure~\ref{fig:F} correspond to a regime where $\mathcal{C}$ is equal to $-1.2$ and $-2.4$, which do not satisfy our requirement for the existence of long distance modes.\footnote{Notice that, being our solution formally identical to the one at $T=0$, the free energy $F=U-TS$ can be negative only if the energy density, and hence $\mathcal C$, are negative as well.}   However, one can compute the behavior of the free energy when $\mathcal{C}$ is changed, as reported in Figure~\ref{fig:FvsC}.

It is clear that when we approach a regime where the UV and IR cutoffs are largely separated the free energy becomes more negative. This means that not only the phase transition will still occur, but the critical temperature will be even higher. Thus, despite the limitations of our numerical computation, the conclusions we reached in Section~\ref{sec:phase transition} still hold.


\bibliographystyle{apsrev4-1}
\bibliography{biblio}

\end{document}